\documentclass[aps,prl,twocolumn,superscriptaddress, floatfix, longbibliography]{revtex4-2}
\usepackage{amsfonts, amssymb, amsmath}
\usepackage{graphicx}
\usepackage{float}
\usepackage[plain]{fancyref}
\usepackage{placeins}
\usepackage{hyperref}
\usepackage{color,soul}
\usepackage{siunitx}
\usepackage{natbib}
\usepackage[version=3]{mhchem} 
\usepackage{physics}
\usepackage{pdfrender}
\newcommand{\gll}{G_\mathrm{LL}}
\newcommand{\glr}{G_\mathrm{LR}}
\newcommand{\grl}{G_\mathrm{RL}}
\newcommand{\grr}{G_\mathrm{RR}}

\begin{document}

\widetext

\title{Nonlocal measurement of quasiparticle charge and energy relaxation in proximitized semiconductor nanowires using quantum dots}

\author{Guanzhong~Wang}
\author{Tom~Dvir}
\email{t.dvir@tudelft.nl}
\author{Nick~van~Loo}
\author{Grzegorz~P.~Mazur}
\affiliation{QuTech and Kavli Institute of NanoScience, Delft University of Technology, 2600 GA Delft, The Netherlands}
\author{Sasa~Gazibegovic}
\author{Ghada~Badawy}
\author{Erik~P.~A.~M.~Bakkers}
\affiliation{Department of Applied Physics, Eindhoven University of Technology, 5600 MB Eindhoven, The Netherlands}
\author{Leo~P.~Kouwenhoven}
\author{Gijs~de~Lange}
\email{gijs.delange@microsoft.com}
\affiliation{Microsoft Quantum Lab Delft, 2600 GA Delft, The Netherlands}

\date{\today}

\begin{abstract}
The lowest-energy excitations of superconductors do not carry an electric charge, as their wave function is equally electron-like and hole-like. 
This fundamental property is not easy to study in electrical measurements that rely on the charge to generate an observable signal. 
The ability of a quantum dot to act as a charge filter enables us to solve this problem and measure the quasiparticle charge in superconducting-semiconducting hybrid nanowire heterostructures.
We report measurements on a three-terminal circuit, in which an injection lead excites a non-equilibrium quasiparticle distribution in the hybrid system, and the electron or hole component of the resulting quasiparticles is detected using a quantum dot as a tunable charge and energy filter. 
The results verify the chargeless nature of the quasiparticles at the gap edge and reveal the complete relaxation of injected charge and energy in a proximitized nanowire, resolving open questions in previous three-terminal experiments.
\end{abstract}

\maketitle

\section*{Introduction}
The elementary excitations in superconductors (SC) are Bogoliubov quasiparticles (QPs), i.e., \ the superposition of an electron excitation with amplitude $u(E)$ and a hole excitation with amplitude $v(E)$, where $E$ is the energy of the excitation. 
The electron and hole components are energy-dependent and are given by $|u(E)|^2 = \frac{1}{2}\left( 1 + \sqrt{E^2 - \Delta_\mathrm{SC}^2}/E \right) $ and $|v(E)|^2 = \frac{1}{2}\left( 1 - \sqrt{E^2 - \Delta_\mathrm{SC}^2}/E \right) $, where $\Delta_\mathrm{SC}$ is the superconducting energy gap, and the QP energy obeys $E>\Delta_\mathrm{SC}$. 
The charge of the excitation, given by $q(E) = e \left( |v(E)|^2 - |u(E)|^2 \right)$, varies between $-e$ (electron-like) when the excitation energy is far above the Fermi energy and $+e$ (hole-like) when it is far below. 
In the vicinity of the gap, $E \approx \Delta_\mathrm{SC}$, the QPs consist of nearly equal superpositions of electron and hole parts as $|u(E)| \approx |v(E)| \approx 1/\sqrt{2}$. 
Therefore, their charge approaches zero \cite{Tinkham1996}.  

In thermal equilibrium and when $k_\mathrm{B} T \ll \Delta_\mathrm{SC}$, where $k_\mathrm{B}$ is the Boltzmann constant and $T$ the electron temperature, the presence of the energy gap ensures that almost no QP excitations exist in the system. 
However, external perturbations such as injection of charge can drive the system out of equilibrium, making its distribution function, $f$, deviate from the Fermi-Dirac distribution, $f_\text{FD}$.
This departure can be decomposed into several non-equilibrium modes \cite{Heikkila2019}.
The two most discussed modes in a superconductor are energy and charge non-equilibrium.
In transport experiments, injecting electrons above the SC gap brings extra charge as well as energy into the system, exciting both modes.
Although the distribution is restored to $f_\text{FD}$ far away from the perturbation, each mode relaxes to the equilibrium over different length and time scales.
This process is studied thoroughly for metallic SCs, both theoretically and experimentally \cite{Clarke1972, Tinkham1972, Tinkham1972b, Kivelson1990, Giazotto2006, MacHon2013, Bergeret2018}.

The discussion so far considered intrinsic superconductors where the energy gap and the electron-hole correlations are generated by an internal pairing mechanism, such as electron-phonon coupling.
Similar effects are also found in SC-proximitized semiconducting (SM) systems, where electron-hole correlations are induced by the proximity effect instead.
Here, similar to the formation of Andreev bound states (ABSs) in a confined system, Andreev reflection (AR) on the SM-SC interface pairs states above and below the Fermi surface in the SM to create a superconducting-like band structure with an induced gap $\Delta_\mathrm{ind}$.
Its relative size $\Delta_\mathrm{ind}/\Delta_\mathrm{SC}$, being between 0 and 1, is directly related to the coupling strength between the SC and SM \cite{antipov_effects_2018}.
These proximitized states are also superpositions of electrons and holes with energy-dependent amplitudes $u(E)$ and $v(E)$, respectively. 
Similar to intrinsic SC, at $E \approx \Delta_\text{ind}$, the electron and hole components are nearly equal, driving the charge of the lowest-energy proximitized states to zero \cite{Bauer2007, Meng2009, Lee2014}. 
This effect was measured both for discrete ABSs \cite{Menard2019} and proximitized semiconducting subbands \cite{Denisov2021}.

Nonlocal transport is a typical experimental tool to study non-equilibrium modes and their relaxation. 
Such a setup utilizes two tunnel junctions: an injector and a detector. 
The injector junction injects particles into the system under study, exciting one or more non-equilibrium modes. 
The detector junction is usually unbiased and measures the response of the system at some distance from the injection point. 
Electron transport is well-suited to measure the charge non-equilibrium mode \cite{Clarke1972, Pothier1997} and can be adapted to measure spin imbalance \cite{Quay2013, Hubler2010}.
However, the energy mode is harder to measure this way, since electron and hole currents flowing into the unbiased probe cancel each other out by virtue of the charge neutrality of this mode \cite{denisov_heat-mode_2020}.
An energy non-equilibrium mode will, however, generate a measurable charge current if the transmission probability of the tunnel barrier, $\mathcal{T}$, is energy-dependent and filters out only one type of carriers \cite{Hussein2019,tan2021}.
A semiconducting quantum dot (QD), with its single-electron orbitals having sharply-peaked transmission amplitudes only for a particular charge, energy and spin, is such a transmission filter.
Previous works have made use of the energy filtering effects of a QD to probe the non-equilibrium distribution of quantum Hall edge states \cite{altimiras2010non}.
As demonstrated later in this text, similarly applying the charge filtering capabilities of QDs opens up new avenues to the study of non-equilibrium in hybrid SM-SC systems.

Nonlocal conductance (NLC) can also serve to measure the sign of the charge of ABSs \cite{Schindele2014, Menard2019} and other effects such as crossed Andreev reflection \cite{Recher2001,Russo2005,Schindele2012,Das2012b,Schindele2014,Tan2015}.
Recently it was further suggested as a powerful tool to measure the induced superconducting gap in semiconducting nanowires \cite{Rosdahl2018}. 
NLC was used to differentiate between bulk induced gap closing from the presence of local resonances in tunnel barriers \cite{Puglia2020,pikulin_protocol_2021}.
All reported measurements in such geometries share common characteristic features---e.g., predominantly anti-symmetric NLC whose global sign is heavily influenced by the tunnel barrier on the current-receiving side.
We show that these effects can arise from the charge and energy dependence of $\mathcal{T}$, which is ubiquitous in gate-defined tunnel barriers.

\bigskip

In this work, we study NLC in a hybrid SC heterostructure. Gate-defined QDs separating the ohmic leads from the hybrid segment are used as charge and energy filters.
We detect independently the electron and hole components of the QP wavefunction, observing charge neutrality of the excitations at the superconducting gap edge. 
Applying a magnetic field that closes the induced gap shows that the charge-to-energy conversion is independent of the presence of an induced spectral gap and only requires Andreev reflection at the SM-SC interface. 
Finally, using QDs to inject and detect QPs, we demonstrate complete relaxation of the detected QPs into the lowest excited states.

\section*{Results}
\subsection*{Qualitative description of the experiment}

\begin{figure*}[ht!]
    \centering
    \includegraphics[width=\textwidth]{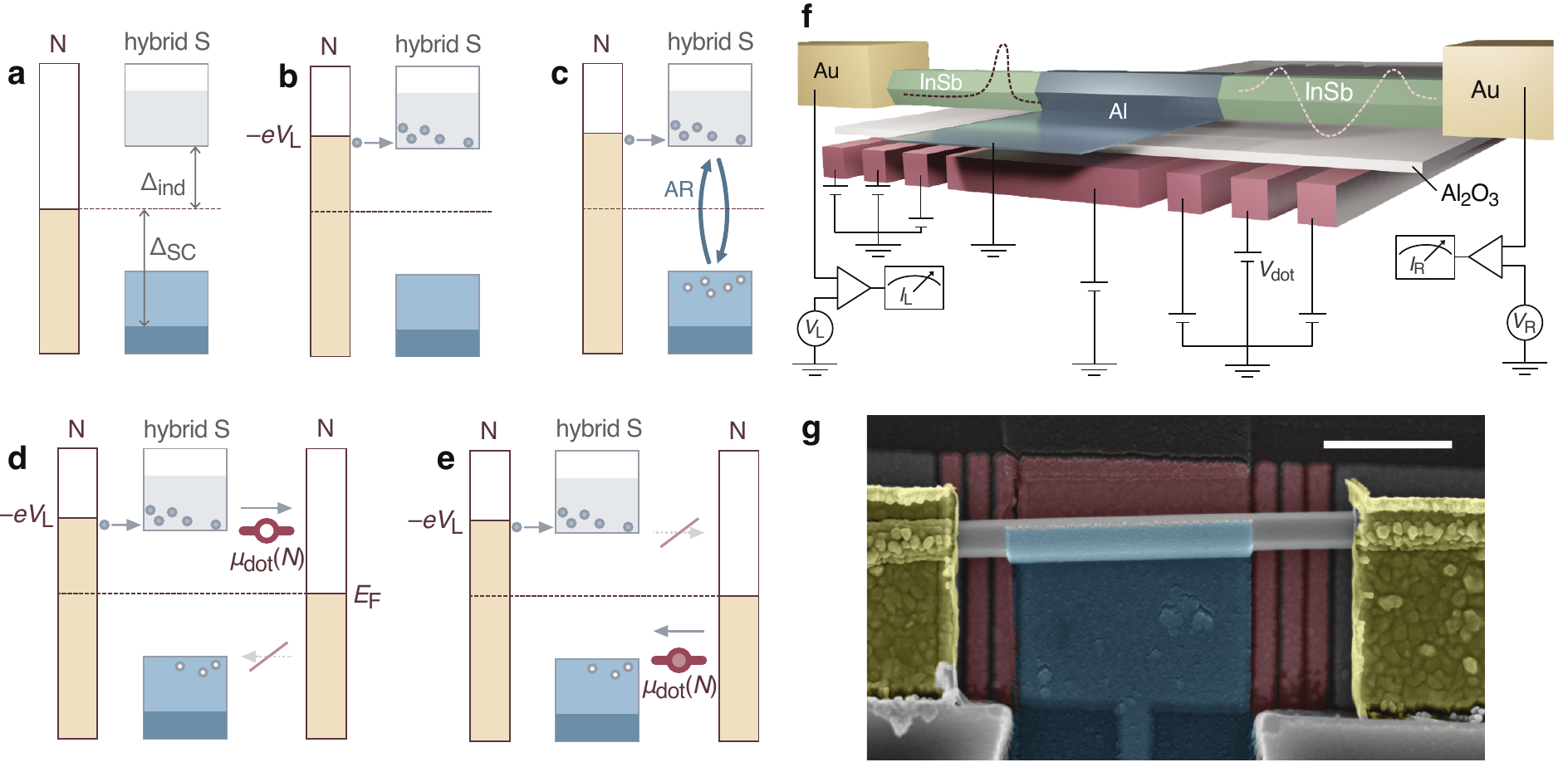}
    \caption{\textbf{a--c.} Illustrations of non-equilibrium modes in proximitized superconductors. 
    a: Thermal equilibrium with no QP present. 
    b: Injection of electrons excites both charge and energy non-equilibrium modes. 
    c: Pure energy non-equilibrium mode after the charge mode has relaxed. The illustrations in this paper are not to scale.
    \textbf{d, e.} A scheme of the nonlocal experiment: a \SI{1}{\micro\meter}-long, grounded superconductor with a gap $\Delta_\text{SC}$ is attached to the middle section of the nanowire, inducing a proximity gap $\Delta_\text{ind}$. 
    On the left side, a tunnel junction under finite bias injects current. 
    On the right side, a quantum dot filters either electrons (d) or holes (e) depending on its occupation. 
    Arrows indicate directions of electron hopping. 
    \textbf{f.} Device schematic and measurement setup. 
    Sketches in the InSb wire segments indicate the potential landscape created by the finger gates. 
    The voltages applied on each group of finger gates are schematically represented by the height of the voltage sources in the circuit diagram. 
    Unlabeled voltages are kept constant within each data set. \textbf{g.} False-colored scanning electron microscopy (SEM) image of a lithographically identical device to Devices~A and B presented in this work. Bottom gate electrodes are colored in red and separated from the InSb nanowire by a thin layer of atomic-layer-deposited (ALD) \ce{Al2O3} dielectric of around \SI{20}{nm} (invisible in this image). The middle section of the nanowire is covered by a thin, superconducting layer of Al film colored in cyan. Two Au ohmic leads are colored in yellow and form the left and right contacts. Scale bar is \SI{500}{nm}. }
    \label{fig: SEM}
\end{figure*}

Figure~\ref{fig: SEM}(a) sketches a hybrid SC heterostructure attached to a normal lead (N) through a tunnel junction.
The proximity effect from the SC opens an induced gap $\Delta_\mathrm{ind}$.
The system is in equilibrium, and if $\Delta_\mathrm{ind} \gg k_\mathrm{B} T$, almost no QPs are excited in the system. 
When we apply a bias $V_\textrm{L}$ such that $-eV_\textrm{L} > \Delta_\textrm{ind}$ [Fig.~\ref{fig: SEM}(b)], electrons are injected into the proximitized semiconductor.
However, the native excitations are not electrons, but Bogoliubov QPs with a charge generally smaller than the elementary charge. Thus, pure electron-like excitations in the injecting lead are eventually converted in the SM-SC hybrid to Bogoliubov quasiparticles, i.e., excitations consisting of a superposition of electron-like and hole-like particles with amplitudes $u(E)$ and $v(E)$, as discussed above.
During this process, the excess charge is drained to the ground through AR at the interface with Al [Fig.~\ref{fig: SEM}(c)].
Until the eventual recombination of all QPs back into the Cooper pair condensate, they carry a finite energy excitation \cite{Bergeret2018,Kivelson1990}.

To measure the generated non-equilibrium distribution, we attach another lead on the right, separated by a semiconducting junction. 
A conventional tunnel junction supports bi-directional currents, where electrons tunnel to unoccupied states above $E_\text{F}$ and holes tunnel to occupied states below it. 
When both charge carriers are present, such as in a SC, the current is proportional to the charge of the QP, $q$ \cite{Danon2020}, which is expected to vanish at the gap edge. 
A QD embedded in a junction can be tuned to only allow current flow in a single direction.
When the charging energy of the QD is far greater than other energy scales in the circuit ($\Delta_\mathrm{ind}$ and bias voltage) and it is tuned to be near a charge degeneracy between having $N$ and $N-1$ electrons, it can be considered a single isolated Fermionic level.
If this level is at $\Delta_\mathrm{ind}>0$ and thus its ground state is empty [Fig.~\ref{fig: SEM}(d)], it allows only electrons to flow from the SM-SC system to the N lead, with a rate proportional to the electron component $|u|^2$ of the proximitized states. 
Similarly, a level at $-\Delta_\mathrm{ind}$, being occupied in its ground state, only allows holes to tunnel to N with a rate proportional to $|v|^2$ [Fig.~\ref{fig: SEM}(e)].
Thus, a QD is a charge-selective probe that couples to either the electron or hole component of the QP wavefunction depending on its occupation.

\subsection*{Methods}

The sample was fabricated using the same methods described in Ref~\cite{Heedt2020}.
A 3D illustration and a scanning electron microscope image of the device geometry are shown in Figure~\ref{fig: SEM}(f,g).
Ti+Pd local bottom gate electrodes were evaporated on Si substrates followed by HSQ shadow wall structures and then atomic-layer deposited \ce{Al2O3} dielectric.
InSb nanowires were grown by MOVPE~\cite{badawy2019}. 
The nanowires were then transferred using an optical manipulator to the substrate described above.
Atomic H cleaning removed the oxide on InSb, and following in-situ transfer in the same e-beam evaporator, \SI{14}{nm} of Al thin film was deposited at liquid \ce{N2} temperature.
The film covers the nanowire middle segment, proximitizing InSb and forming our galvanically connected middle lead (S).
Finally, the N leads were fabricated by another e-beam lithography step.
After patterning, Ar milling removes the newly formed surface oxide again, and \SI{140}{nm} of Cr+Au contact was evaporated to form ohmic contacts to InSb. The hybrid SC-SM segment in the devices reported here is \SI{1}{\micro m} long, and the typical distance between the centers of the finger gates forming the QD is \SI{110}{nm}.
Overall, 5 devices showing the same qualitative behavior were measured. 
Here we report on detailed scans of two such devices.

 \begin{figure}[h!]
    \centering
    \includegraphics[width=0.48\textwidth]{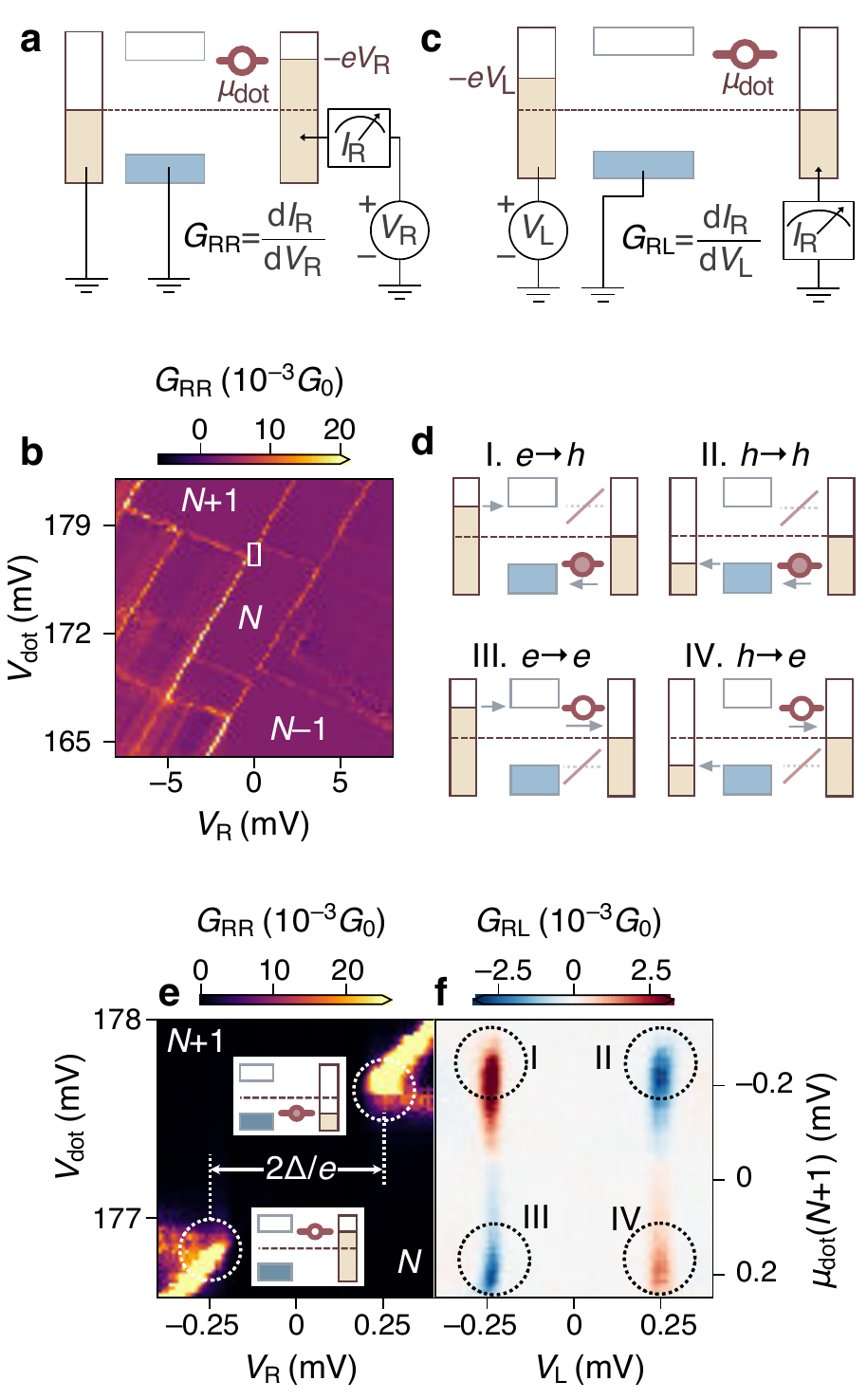}
    \caption{
    \textbf{a.} Circuit diagram for local conductance $\grr$ measurement of Device~A. 
    \textbf{b.} $\grr$ through the right N-QD-S junction.
    Charge occupations numbers are labeled within corresponding Coulomb diamonds, $N$ being odd.
    The white box indicates the zoomed-in area in e.
    \textbf{c.} Circuit diagram for nonlocal conductance $\grl$ measurements.
    \textbf{d.} Energy alignment between the QD and leads for the four injected vs detected charge type possibilities when measuring $\grl$.
    Directions of electron tunneling are indicated with arrows.
    \textbf{e.} Higher-resolution scan of the local conductance of the right side $\grr$ near the QD charge degeneracy point marked in panel~b. 
    Insets: sketches of the energy alignment between the S, right N, and the QD level at the Coulomb-diamond tips (circled in white). 
    \textbf{f.} The NLC $\grl$ measured on the right side as a function of the bias on the left side $V_\mathrm{L}$ and the right dot gate $V_\text{dot}$, with $V_\text{R} = 0$. 
    The correspondence between the four resonant features and the four situations depicted in panel~d are circled and labeled with roman numerals.}
    \label{fig: Butterfly}
\end{figure}

\subsection*{Detection of quasiparticle charge}

First, we characterize the QD defined using the three finger gates in the right junction.
Figure~\ref{fig: Butterfly}(a) illustrates the circuit used for measurement of the local conductance $\grr \equiv \mathrm{d}I_\mathrm{R} / \mathrm{d}V_\mathrm{R}$ through the QD when applying $V_\mathrm{R}$ and varying $V_\mathrm{dot}$ while keeping S and left N grounded. 
We observe Coulomb diamonds of varying sizes, typical of few-electron QDs [Fig.~\ref{fig: Butterfly}(b)].
By comparing the resonance lines with the constant interaction model of a QD \cite{Kouwenhoven2001}, we extract the capacitance parameters: gate and bias lever arms and the charging energy, thus mapping the applied gate voltage to the QD's chemical potential $\mu_\text{dot}(N) =-e\alpha \left( V_\text{dot} - V_\text{res,N} \right)$ where $\alpha$ is the gate lever arm and $V_\text{res,N}$ is the gate voltage when the $N$ and $N-1$ charge states are degenerate \footnote{See supplementary Figure~S1 for details}. 

The circuit used to measure the nonlocal conductance $\grl \equiv \mathrm{d}I_\mathrm{R} / \mathrm{d}V_\mathrm{L}$ is sketched in Figure~\ref{fig: Butterfly}(c).
The sign of the injecting bias determines the type of charge carriers injected into the hybrid S, electrons if $V_\mathrm{L}<0$ and holes vice versa.
The detected charge, as explained above, is determined by the receiving QD's chemical potential $\mu_\mathrm{dot}(N+1)$ being above or below 0.
Figure~\ref{fig: Butterfly}(d) illustrates the four possible injection versus detection charge combinations when both the left N and QD are on resonance with the SC gap edge.

Zooming in to the $N\rightarrow N+1$ charge transition [Fig.~\ref{fig: Butterfly}(b)], we measure the detailed local and nonlocal conductance structures [Fig.~\ref{fig: Butterfly}(e),~(f), respectively].
The local conductance [Fig.~\ref{fig: Butterfly}(e)] shows QD diamond lines with the exception that transport is blocked at energies smaller than $\Delta_\mathrm{ind} \approx \SI{250}{\micro V}$ \cite{gramich2016subgap, Schindele2014, Lee2014, deacon2010, pillet2013, kumar2014, jellinggaard2016, junger_spectroscopy_2019, junger2020, Bruhat2016,Devidas2021}. 
We note that due to the strong SM-SC coupling $\Delta_\mathrm{ind} \approx \Delta_\mathrm{SC}$ \cite{antipov_effects_2018}.
QDs coupled to SCs can form ABSs\cite{Lee2014, Su2018}.
Such formation requires two-electron tunneling processes to take place between the QD and the proximitized segment.
By raising the tunnel barriers, we significantly suppress such two-electron processes and inhibit the formation of ABSs in the QD. 
The lack of sub-gap features confirms that the QD is not hybridized by the SC and therefore maintains its pure electron or hole character.

In Fig.~\ref{fig: Butterfly}(f), we vary $V_\mathrm{dot}$ through the same resonances while scanning $V_\mathrm{L}$.
At sub-gap energies ($|eV_\mathrm{L}| < \Delta_\mathrm{ind}$), $\grl$ is similarly 0 due to the absence of sub-gap excitations.
$\grl$ is also weak when $V_\text{dot}$ is far away from the tips of the Coulomb diamonds~[Fig.~\ref{fig: Butterfly}(b)]. 
Finite NLC is only obtained when the left bias is aligned with the induced SC gap edges, $eV_\text{L}\approx \pm \Delta_\mathrm{ind}$, and the QD level is inside the induced gap: $-\Delta_\mathrm{ind} \le \mu_\text{dot}(N+1) \le \Delta_\mathrm{ind} $.
The NLC feature around the QD crossing contains four lobes that exhibit a two-fold anti-symmetry, changing signs under either opposite bias or dot occupation.
This four-lobed structure corresponds exactly to the four charge combinations in Figure~\ref{fig: Butterfly}(d) and shows up in almost all charge degeneracy points we have measured, including other dot configurations and devices \footnote{see Supplemental Figure~S2 for other devices showing the same structure}.

Consider the feature marked by ``I" in Figure~\ref{fig: Butterfly}(f) and the process schematically depicted [Fig.~\ref{fig: Butterfly}(d)]. 
The electrons injected by the negative left bias into the central region create both energy and charge non-equilibrium. 
The resulting QPs arriving at the right end of the hybrid S must then reach the right N lead via a QD tuned to $\mu_\text{dot}(N+1) \approx -\Delta_\mathrm{ind}$, which only allows holes to tunnel out.
The presence of this NLC lobe is thus a result of the hole component of the wavefunction $v$. 
The inversion of charge in this $e\rightarrow h$ process results in the observed positive NLC due to the current-direction convention.
We can also tune the dot to $\mu_\text{dot}(N) \approx +\Delta_\mathrm{ind}$ (marked ``III"). 
Here electrons are still injected into the central region, but now the QD allows only electrons to tunnel out to the right lead because its ground state is an unoccupied fermionic level. 
We mark this process $e \rightarrow e$.
The NLC is thus negative, with a magnitude that relates to $u$.

The NLC is also predominantly anti-symmetric with respect to the applied voltage bias. 
This can be understood by considering the current-rectifying behavior of the QD at a fixed occupancy. 
When $\mu_\text{dot}>0$, only the flow of electrons from the S to the N lead is allowed, regardless of the charge of the injected particles. The current passing through the dot is thus always positive, forcing the conductance ($\textrm{d}I/\textrm{d}V$) to flip its sign when the bias changes polarity. 
Similarly, when the dot is placed at $\mu_\text{dot}<0$ to allow only holes to flow, the current is always negative, and the rest follows suit.
Anti-symmetric NLC is also a prevalent feature in conventional tunnel junction measurements without any QD \cite{Puglia2020}. 
There, similar to our observations, the global sign of the anti-symmetric NLC is determined by and varies with the gate on the current-detecting junction. 
We argue that this ubiquitous anti-symmetry with respect to bias voltage stems from the unintentional charge selectivity of the semiconducting tunnel junctions~\footnote{See Supplemental Figure~S3 for data supporting this interpretation, when both sides of the device are configured to be tunnel junctions}.

We note that the amplitudes of the $\grl$ peaks are higher when the ground state contains $N+1$ electrons than when it contains $N$ electrons [Fig.~\ref{fig: Butterfly}(f)], with $N$ being odd in this setup.
This difference in $\grl$ can be attributed to the spin-degenerate DOS of the dot, which gives rise to different tunneling rates for even and odd occupation numbers~\cite{Johnson1992, VanVeen2018}. 
The opposite trend can be observed in the $N-1\rightarrow N$ transition and the application of a small Zeeman field that lifts this degeneracy restores the electron and hole's amplitudes to be nearly equal~\footnote{see Supplemental Figure~S4 for the effect of lifting Kramer's degeneracy on the NLC amplitude}. 
Spin degeneracy influences the QD transmission rates of electrons and holes in a manner unrelated to the relative strengths of $u$ and $v$, thus obscuring the observation of charge neutrality. 

\subsection{Many-electron dots}

\begin{figure}[ht]
    \centering
    \includegraphics[width=0.5\textwidth]{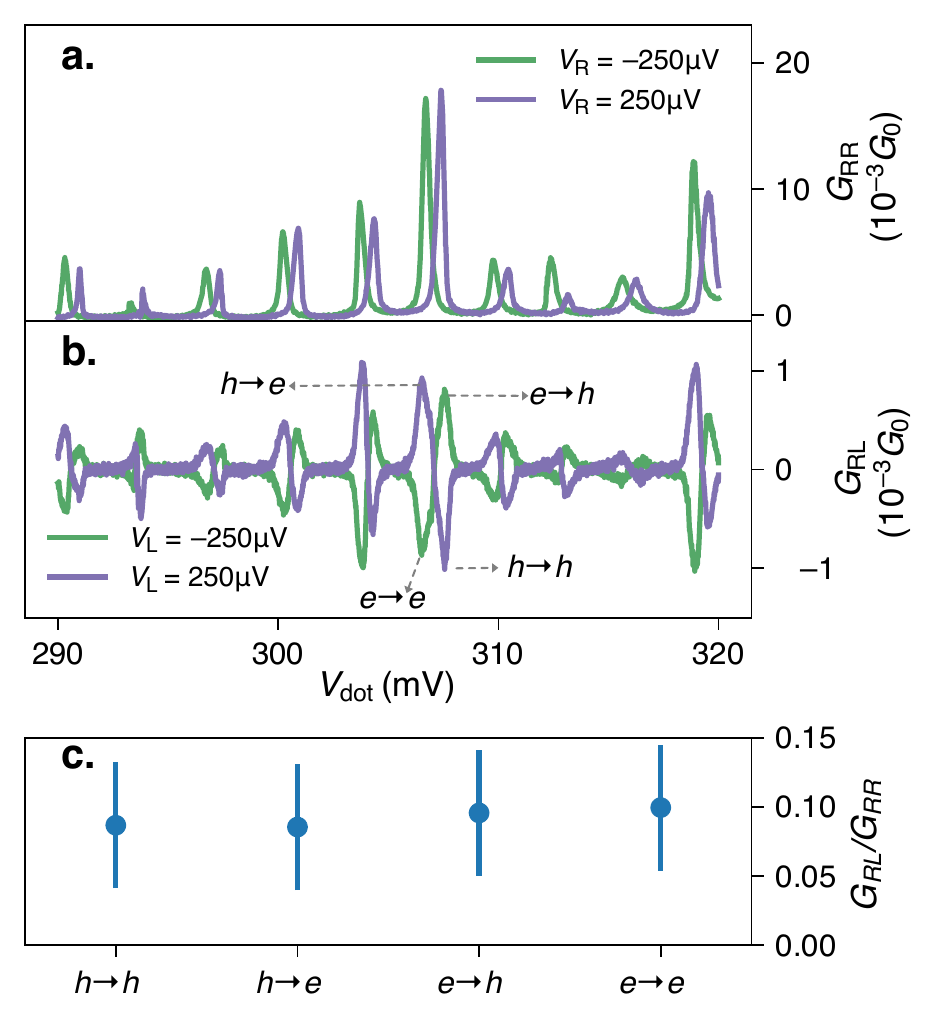}
    \caption{ \textbf{a} The local conductance $\grr$ of Device~B as a function of $V_\mathrm{dot}$, measured with fixed $V_\mathrm{R} = \pm \SI{250}{\micro V}$ . 
    \textbf{b.} The nonlocal conductance $\grl$ of Device~B as a function of $V_\mathrm{dot}$, measured with fixed $V_\mathrm{L} = \pm \SI{250}{\micro V}$, respectively.
    The four peaks in one of the periodic structures are labeled by their injected and detected charge states being electrons ($e$) or holes ($h$). 
    \textbf{c.} The relative magnitude of $h \rightarrow h$, $h \rightarrow e$,  $e \rightarrow h$, and $e \rightarrow e$ processes to the magnitude of the local conductance through the QD in the same configuration. 
    See details in text.
    }
    \label{fig: Many}
\end{figure}

The few-electron QD discussed above proves to be an effective charge filter, able to detect separately the electron and hole components of the QP wavefunction. 
However, the presence of spin degeneracy modifies the tunneling rates of electrons and holes and complicates direct comparison between $u$ and $v$. 
To overcome this, we turn our attention to Device~B, which has a larger QD whose orbital level spacing is too small to be observed.
Here, since multiple orbital levels contribute to tunneling across, rendering dot and the spin degeneracy negligible, the tunneling rates for electrons and holes are nearly equal.
Figure~3(a) shows the local conductance $\grr$ through the QD as a function of the gate voltage, when applying a bias of $V_\textrm{R} =  \pm \Delta_\mathrm{ind}/e = \pm\SI{250}{\micro V}$ between the N and S leads on either side of the QD. 
We observe equidistant Coulomb oscillations typical of many-electron QDs.
The magnitude of the oscillations varies from peak to peak and between positive and negative applied $V_\textrm{R}$, indicating the mesoscopic details of transport are different under device voltage changes.
We expect such differences to modulate the NLC as well. 

The NLC oscillates as a function of $V_\textrm{dot}$ when applying $V_\textrm{L} =  \pm \Delta_\mathrm{ind}/e = \pm\SI{250}{\micro V}$ [Fig.~\ref{fig: Many}(b)]. 
Every period of the oscillation has an internal structure where a positive peak follows a negative peak (green curve) or the opposite (purple curve). 
Each peak in the NLC trace represents a different process in which either electrons ($e$) or holes ($h$) are injected and either the electron ($e$) or hole ($h$) component of the wavefunction is detected. 
With a negative applied bias, electrons are injected into the system.
The negative conductance peak appears first, resulting from the QD detecting the electron component of the wavefunction  ($e \rightarrow e$).
After $\mu_\textrm{dot}$ crossed zero, the QD detects the hole component ($e \rightarrow h$), giving rise to positive conductance.
The choice of injected charge doesn't affect the charge selectivity of our probe. Thus, injecting holes instead of the electrons (positive $V_\text{L}$) leads to an inversion of the sign of $G_\text{RL}$.

In Figure~\ref{fig: Many}(c) we show the amplitude of $\grl$ relative to the amplitude of $\grr$ through the QD, as a function of the four processes, averaged among the multiple resonances shown in Figure~\ref{fig: Many}(a,b).
The average relative amplitudes of all the processes are around 0.1. 
The differences between amplitudes of the four processes are much smaller than variations within each process. 
This is consistent with the BCS picture discussed above, in which the excited states in a superconductor at the gap edge are chargeless Bogoliubov quasiparticles, i.e., $|u| = |v|$. 

\begin{figure}[ht!]
    \centering
    \includegraphics[width=0.5\textwidth]{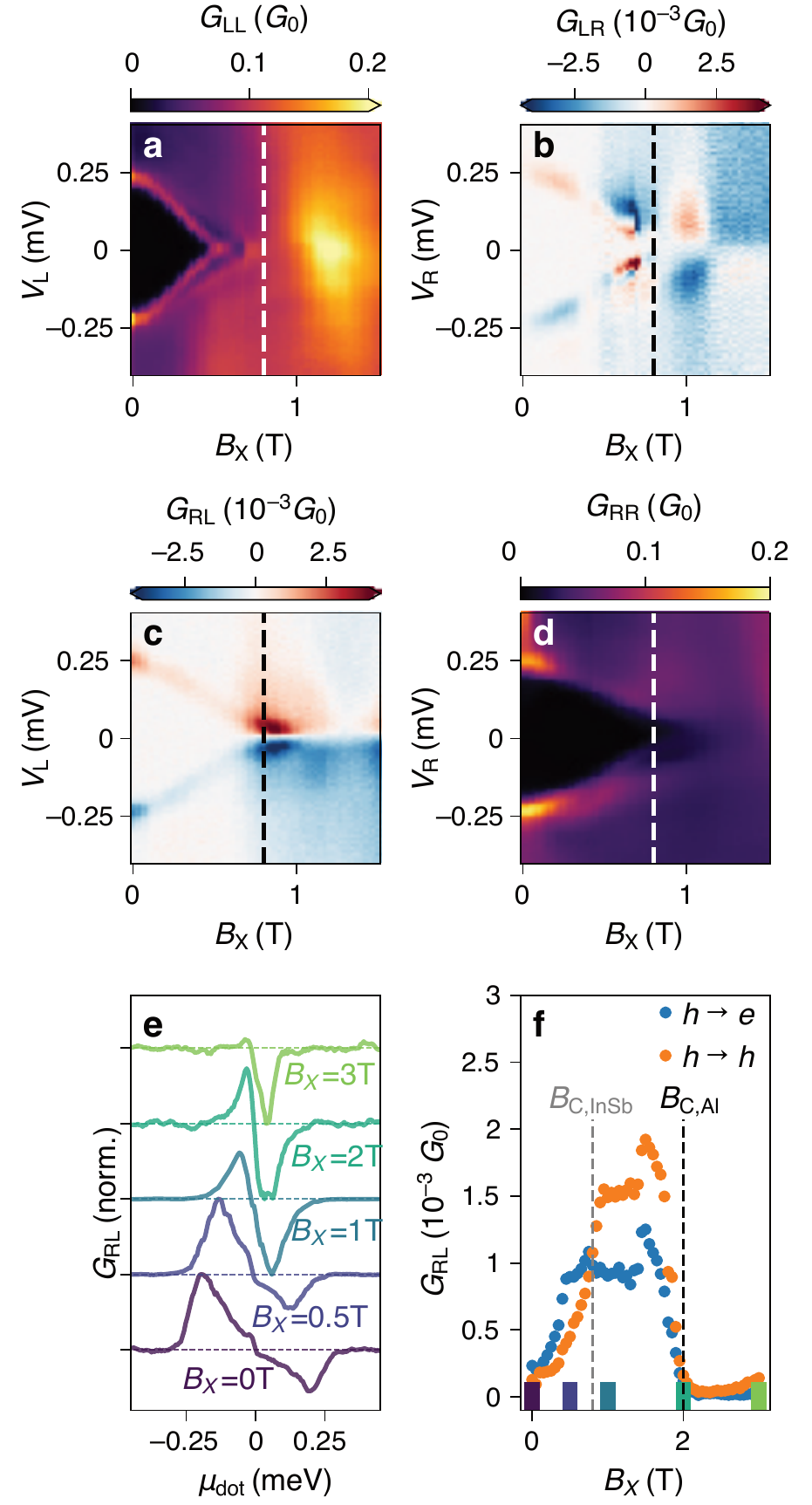}
    \caption{ 
    \textbf{a.-d.} Conductance matrix $G$ as a function of in-plane magnetic field ($B_\mathrm{X}$) and bias.  $\gll$ (a.) and $\grl$ (c.) are measured as a function of $V_\mathrm{L}$. $\grr$ (b.) and $\glr$ (d.) are measured as a function of $V_\mathrm{R}$.
   \textbf{e.} $\grl$ as a function of $\mu_\mathrm{dot}(N)$ at different values of $B_X$ measured in Device~A. 
    The line traces are normalized to the maximum value of each: $G_{RL} (B_X) / \text{max}\left( |\grl (B_X)| \right) $. 
    \textbf{f.} Amplitude of the NLC for the $h \rightarrow e$ (positive NLC, blue) and $h \rightarrow h$ (negative NLC, orange) as a function $B_X$. 
    The colored markers show the values of $B_X$ where the line scans in panel~a are taken.
    The $B$ field at which the spectral gap in InSb closes, $B_\mathrm{C,InSb}$ and at which the superconductivity in Al vanishes, $B_\mathrm{C,Al}$, 
    are indicated by corresponding vertical lines.
    } 
    \label{fig: closing}
\end{figure}

\subsection*{Presence of particle-hole correlation after the closing of the induced gap}
The three-terminal setup we employ allows us to measure the induced gap independently from possible localized ABSs in the vicinity of the junction. 
The topological phase transition predicted to take place in this system would manifest as the closing and re-opening of the induced gap in response to the application of a magnetic field or a change in the chemical potential at a finite magnetic field \cite{Rosdahl2018}. 
In Fig~\ref{fig: closing}a-d, we show the full conductance matrix  $\boldsymbol{G} = \begin{pmatrix} G_\text{LL} & G_\text{LR} \\ G_\text{RL} & G_\text{RR} \end{pmatrix}$ as a function of a magnetic Zeeman field $B_X$ applied along the nanowire direction. 
While the local conductance on the left side, $G_\text{LL}$, shows a local sub-gap state crossing zero bias at a finite field, $G_\text{RR}$ lacks such features. 
The bulk induced gap, seen in the nonlocal conductance, close at $B = \SI{0.8}{T}$ without reopening. 
The lack of gap reopening is consistent with past measurements of the nonlocal conductance in three-terminal geometry \cite{Puglia2020} and most likely results from the presence of disorder as the dominating energy scale in these nanowires \cite{Pan2021}.  

In a conventional superconductor, the presence of chargeless excitations and the existence of an energy gap are correlated (except for a small region in the phase space characterized by gapless superconductivity \cite{Tinkham1996}). 
In a proximitized system, both effects arise separately from Andreev reflection at the SM-SC interface. 
To see this, we measure the nonlocal conductance using a detector QD at different values of $B_X$ [Fig.~\ref{fig: closing}(e)]. 
Since the size of the induced gap decreases upon increasing magnetic field, we apply a constant DC bias of \SI{200}{\micro V} and AC excitation of \SI{180}{\micro V} RMS to ensure the induced gap edge always lies within the measured bias window.
All of the nonlocal scans taken with $B_X<\SI{2.5}{T}$ show a positive and a negative peak arising from the $h \rightarrow e$ and $h \rightarrow h$ process, respectively, as discussed above. 
Taking the maximal positive value of $G_\mathrm{RL}(B_X)$ as the amplitude of the $h \rightarrow e$ process, and the maximal negative value as the amplitude of the $h \rightarrow h$ process, we track the evolution of both as a function of $B_X$ [Fig.~\ref{fig: closing}(f)]. 
Both processes survive well above the gap closing field of \SI{0.8}{T}. 
Only above \SI{2}{T}, the critical field of the Al film, we observe a decay in the NLC amplitude. 
At higher fields the $h \rightarrow e$ process that must involve superconductivity is absent, and the remainder of the $h \rightarrow h$ process may be attributed to voltage-divider effects \cite{Martinez2021}.
The observation of positive nonlocal conductance up to \SI{2}{T} shows that electron-hole correlations persist as long as Andreev reflection between the wire and superconducting film is possible. 
This effect is independent of the presence of an induced gap in the DOS of the proximitized system.

\subsection{Detecting energy relaxation using QDs} \label{sec:energy}

\begin{figure}[ht!]
    \centering
    \includegraphics[width=0.5\textwidth]{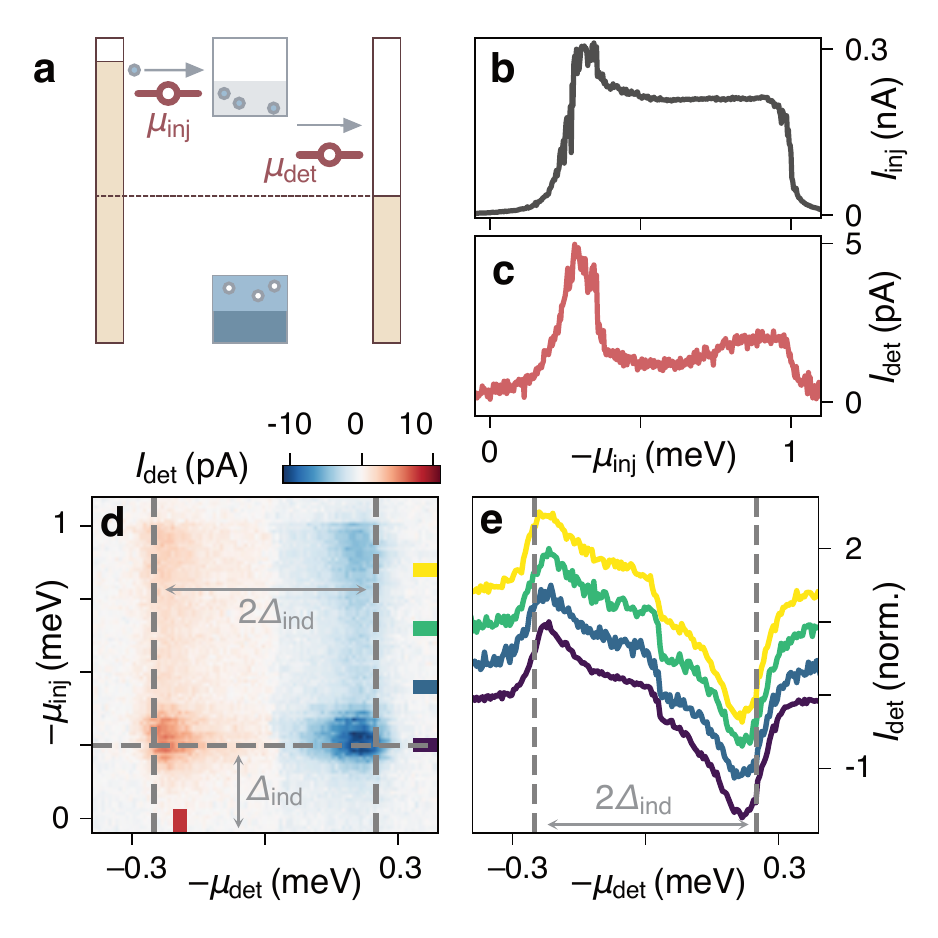}
    \caption{
    \textbf{a. } A sketch of the energy relaxation experiment when the device is operated under negative bias voltage on the left side. (The situation of a positive bias voltage, as the case with the data shown in the following panels, can be similarly represented using a vertically mirrored sketch.) Dots are formed on both sides of the nanowire. The bias is fixed on the right side, and injection energy is determined by the energy of the left dot. The nonlocal current is collected by the right dot. The applied bias is fixed to be  $V_\text{inj}$ = \SI{1}{mV}.
    \textbf{b., c.} Dependence of $I_\text{inj}$ (b.) and $I_\text{det}$ (c.) vs.\ $\mu_\text{inj}$ for fixed $\mu_\text{det} \approx \SI{-240}{\micro V}$ (see corresponding color in panel~d).
    \textbf{d.} $I_\text{det}$ as functions of the chemical potentials of the QDs on the left  $\mu_\text{inj}$ and right  $\mu_\text{det}$.
    The red vertical bar marks the gap edge and detector QD voltage under which curves in panels~b,c are taken.
    Horizontal bars mark the line cuts shown in panel~e.
    \textbf{e. } Normalized line scans of $I_\text{det}$ vs.\ $\mu_\text{det}$ for different injection energies (see corresponding color in panel~d). 
    Lines are vertically offset for clarity.}
    \label{fig: Energy}
\end{figure}

A commonly observed feature in the three-terminal setup we study is the presence of significant NLC when applying a bias greater than the gap of the parent SC, $|eV|>\Delta_\text{SC}$~\cite{Puglia2020,Menard2019}. 
This is unexpected since the injected particles can directly drain to the ground via the available states in the Al film at such energies, and thus should not be able to emerge at the other side of a long enough nanowire.
To study the behavior of the injected QPs at high energies, we tune the finger gates in Device~A to transform the current-injecting junction from a tunnel barrier into a second QD, so that QPs are injected only at controlled energies into the SC [Fig.~\ref{fig: Energy}(a)]. 
We fix the bias on the injecting side to $V_\text{inj} \equiv V_\text{L}=\SI{1}{mV}$ and vary the energy of the injected electrons by scanning the potential of the injecting dot $\mu_\text{inj}$. 
We verify that both inelastic tunneling and elastic co-tunneling across the injector dot are negligible in our device \footnote{see Supplemental Figures~S5, S6 for discussions on the effects of inelastic tunneling and elastic co-tunneling}. 
Thus, the QDs are operated not only as charge filters, but also as energy filters for both injected and detected electrons. 

Next, we show the simultaneously measured currents through the injector dot ($I_\text{inj}$) and the detector dot ($I_\text{det}$) as a function of the injection filter energy $\mu_\text{inj}$ [Fig.~\ref{fig: Energy}(b,c)], when the detector dot is fixed to the gap edge. 
The local current, $I_\text{inj}$, is higher at $\mu_\text{inj} \approx \Delta_\mathrm{ind}$ and decays to a plateau, a result of the high DOS at the gap edge combined with the tunneling out of the QD being primarily elastic.
$I_\text{inj}$ depends only on $\mu_\text{inj}$, showing that coherent effects such as crossed Andreev reflection are negligible \cite{Recher2001,Schindele2014,Schindele2012,Russo2005,Das2012b,Tan2015} \footnote{see the dependence of $I_\text{inj}$ on $\mu_\text{det}$ in Supplementary Figure~S7}. 
The nonlocal current [$I_\text{det}$, Fig.~\ref{fig: Energy}(c)] is smaller than $I_\text{inj}$ by two orders of magnitude. 
It also has a peak when current is injected directly at the gap edge, but persists when injection energy is higher. 

The nonlocal current depends strongly on both $\mu_\text{inj}$ and $\mu_\text{det}$ [Fig.~\ref{fig: Energy}(d)].
Remarkably, we see that while the injection energy modulates the magnitude of the detector current, it does not influence the energy range in which finite $I_\text{det}$ can be measured. 
Fig~\ref{fig: Energy}(e) shows $I_\text{det}$ vs.\ $\mu_\text{det}$ for different values of $\mu_\text{inj}$, normalized by their maximal values. 
They all follow the same trend---regardless of the injected electrons' energy, the nonlocal signal is only collected at energies comparable to or smaller than $|\Delta_\mathrm{ind}|$.
The observation that electrons injected at energies larger than $\Delta_\mathrm{ind}$ are detected only at the gap edge implies inelastic relaxation plays an important role in nonlocal conductance. 
Electrons injected above the gap are either drained by the grounded SC or decay to the gap edge, after which they are free to diffuse and reach the detector junction.
This observation explains why we do not observe a finite QP charge even though they are only neutral at the gap edge while the QDs have a finite energy broadening --- as long as the linewidth of the QD is smaller than $\Delta_\mathrm{ind}$, the only QPs available for transport are those with energy $\Delta_\mathrm{ind}$.

\section*{Discussion}
Summing up our observations of charge and energy relaxation above, the emerging microscopic picture of nonlocal charge transport in three-terminal nanowire devices thus consists of four (possibly simultaneous) processes. 
First, a charge is injected at some given energy into the nanowire.
Second, some of the injected electrons/holes are drained to the ground via the superconducting lead and the remaining relax to the lowest available state at the induced gap edge. 
Third, through Andreev reflection, the charged electrons are converted to chargeless Bogoliubov QPs. 
Finally, the QPs diffuse toward the other exit lead, where they are projected onto a charge polarity determined by the receiving junction. 

In contrast to the QP continuum investigated here, previous works have examined NLC produced by transport through discrete sub-gap Andreev bound states in similar N-S-N hybrid devices~\cite{Menard2019,Danon2020}.
Distinctly from chargeless Bogoliubov QPs at the gap edge, the BCS charge of these sub-gap states is in general nonzero and varies drastically with gates.
These works concluded that the NLC produced by such states is proportional to their BCS charge, implying $\glr=\grl=0$ when $|u|=|v|$.
However, as observed here and in accordance with other works detecting NLC of the QP continuum~\cite{Puglia2020}, anti-symmetric NLC is still consistently detected even at the gap edge where BCS charge of the QPs is expected to vanish.
We argue that finite NLC of chargeless QPs stems from non-ideal tunnel barriers that transmit electrons and holes with non-equal, energy-dependent probabilities, a generic property of semiconducting junctions. 
As the N-S tunnel junctions also exhibit a preference for a certain charge (see Supplemental Figure~S3), even when $|u|=|v|$, the tunnel barrier transmits more electrons than holes.
Thus, the resulting NLC, being proportional to the difference between these two transmission amplitudes~\cite{Danon2020}, becomes finite and antisymmetric. 
To put the above into the scattering formalism, the commonly used framework to describe electron transport in this system, the NLC is given by:
\begin{equation}
    G_{ij}(E) = \frac{e^2}{h}\left(T^{eh}_{ij} - T^{ee}_{ij} \right)
\end{equation}
where $T^{ee}_{ij}$ and $T^{eh}_{ij}$ are the energy-dependent transmission amplitudes from an electron in lead $i$ to an electron or a hole in lead $j\ne i$ \cite{Rosdahl2018}.  The presence of charge and energy filters, such as the QDs employed in this work, significantly modifies these transmission amplitudes. When the QD's chemical potential is tuned above the Fermi level, $T^{eh}_{ij}$ is suppressed, whereas when the QD's chemical potential is tuned below the Fermi level, $T^{ee}_{ij}$ is suppressed. 
We further note that although one is tempted to associate the NLC observed in our system directly with the relevant transmission amplitudes expressed here, the scattering formalism itself is insufficient to model some important aspects of the actual transport. 
The scattering formalism assumes that the motion of the QPs in the system conserves energy \cite{Maiani:2022}, but this assumption does not hold in our system since we observe that energy relaxation plays an important role in NLC. 
We thus conclude that while the present or absence of NLC can serve as a useful tool in the determination of the induced gap, a quantitative analysis of NLC should go beyond the scattering formalism.

\section*{Conclusions}

In conclusion, by measuring the nonlocal conductance in a three-terminal device with well controlled QDs at the ends, we can detect the electron and hole components of non-equilibrium quasiparticle wavefunctions. 
Our results reveal a population of fully charge-relaxed neutral BCS excitations at the gap edge in a proximitized nanowire under nonlocal charge injection.
We further show that the conversion of injected charge to correlated electron-hole excitations does not require an induced gap.
By injecting particles at specific energies, we observe the inelastic decay of injected charges to the lowest excited states, the gap edge.
We show that the combined effect of charge neutral excitation and a tunnel barrier with energy selectivity leads to a current-rectifying effect. 
These results allow us to understand the ubiquitous anti-symmetry of the nonlocal conductance observed in previous reports \cite{Puglia2020} and suggest that the correct framework to discuss such experiments is in terms of non-equilibrium superconductivity.
Crucially, we show that inelastic decay of injected quasiparticles dominate nonlocal transport and therefore should be taken into account when attempting to model the system.

The results observed here are in very good agreement with the results obtained by Denisov et al.~\cite{Denisov2021}, measuring the nonlocal response of Al-covered InAs nanowires. 
There, the problem of detecting the charge-neutral mode was resolved by measuring the nonlocal shot noise response, showing the charge neutrality of the excitations within the hybrid nanowire.
The alternative approach presented here, utilizing QDs as as energy- and charge-selective injectors and detectors supplements the shot noise measurements by uncovering the relaxation processes taking place in the transport process. It can further allow the study of non-equilibrium distribution functions in proximitized semiconducting systems with spectral resolution. In the presence of discrete ABSs or QPs occupying a wide energy range, the energy resolution of a QD allows one to excite and probe different energies as desired.
We further propose that in the presence of magnetic fields, QDs can also serve as efficient bipolar spin filters \cite{hanson_semiconductor_2004}, allowing us to directly measure the spin-polarized density of states of proximitized SC, triplet SC correlations, and extract the relevant relaxation rates and mechanisms. 

\bigskip

\begin{acknowledgments}
We wish to acknowledge useful discussions with Marco Aprili, Torsten Karzig, Jelena Klinovaja, Daniel Loss, Filip Malinowski, Dimitry Pikulin, Charis Quay, Hadar Steinberg and Bernard van Heck.
This work has been supported by the Dutch Organization for Scientific Research (NWO), the Foundation for Fundamental Research on Matter (FOM) and Microsoft Corporation Station Q. 

G. W.\ and T. D.\ contributed equally to this work.

All raw data in the publication and the analysis code used to generate figures are available at \url{https://zenodo.org/record/5534254}.

\end{acknowledgments}

\bibliography{bibliography.bib}

\end{document}


\widetext

\title{Nonlocal measurement of quasiparticle charge and energy relaxation in proximitized semiconductor nanowires using quantum dots \\ Supplementary information}
\author{Guanzhong~Wang}
\author{Tom~Dvir}
\email{t.dvir@tudelft.nl}
\author{Nick~van~Loo}
\author{Grzegorz~P.~Mazur}
\affiliation{QuTech and Kavli Institute of NanoScience, Delft University of Technology, 2600 GA Delft, The Netherlands}
\author{Sasa~Gazibegovic}
\author{Ghada~Badawy}
\author{Erik~P.~A.~M.~Bakkers}
\affiliation{Department of Applied Physics, Eindhoven University of Technology, 5600 MB Eindhoven, The Netherlands}
\author{Leo~P.~Kouwenhoven}
\author{Gijs~de~Lange}
\email{gijs.delange@microsoft.com}
\affiliation{Microsoft Quantum Lab Delft, 2600 GA Delft, The Netherlands}
\date{\today}
\maketitle

\begin{figure}[h!]
    \centering
    \includegraphics[width=0.9\textwidth]{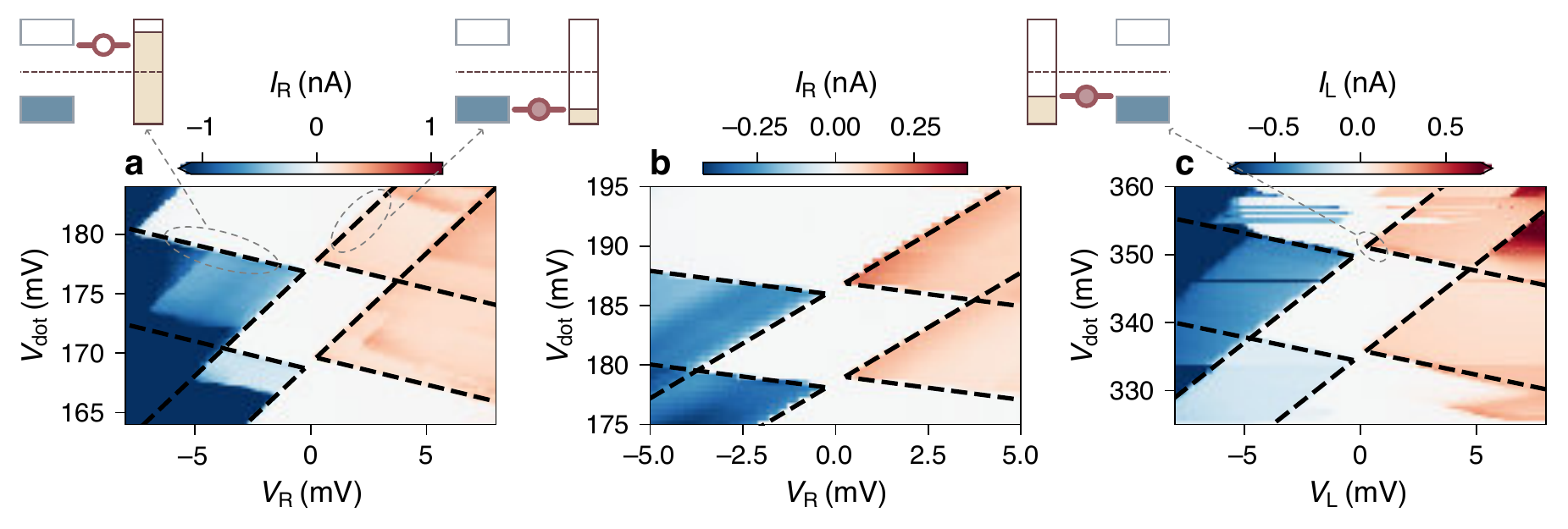}
    \caption{\textbf{Local Coulomb diamonds}
    Conversion between the chemical potential of a quantum dot ($\mu_\mathrm{dot}$) and the applied voltage on the dot's gate ($V_\mathrm{dot}$) requires precise knowledge of the capacitive couplings of the gates to the device, i.e., \ the lever arms. 
    To find this, we measure the current flowing through the dot as a function of the locally applied bias and dot plunger gate voltage. 
    Comparing the Coulomb diamond structures to these measurements, we obtained a lever arm  $\alpha = C_\text{G} / C_\text{tot}$, where $C_\text{G}$ is capacitance between the dot and its plunger gate and $C_\text{tot}$ is the total capacitance of the dot,  of $0.43 \pm 0.01$ and $0.44\pm 0.01$ for the left dot (for two different cutter gate configurations) and $0.29 \pm 0.01$ for the right dot.
    \textbf{a.} $I_\mathrm{R}$ as a function of $V_\mathrm{dot}$ and  $V_\mathrm{R}$, taken with the gate configurations used in Figure~2 of the main text. 
    \textbf{b.} $I_\mathrm{R}$ as a function of  $V_\mathrm{dot}$ and  $V_\mathrm{R}$, taken with the gate configurations used in Figure~4 and Figure~5 of the main text. 
    \textbf{c.} $I_\mathrm{L}$ as a function of  $V_\mathrm{dot}$ and  $V_\mathrm{L}$, taken with the gate configurations used in Figure~5 of the main text. 
    Energy alignments between leads are sketched for a few representative features in the data.}
    \label{fig: dot-level-char}
\end{figure}

\begin{figure}
    \centering
    \includegraphics[width=0.9\textwidth]{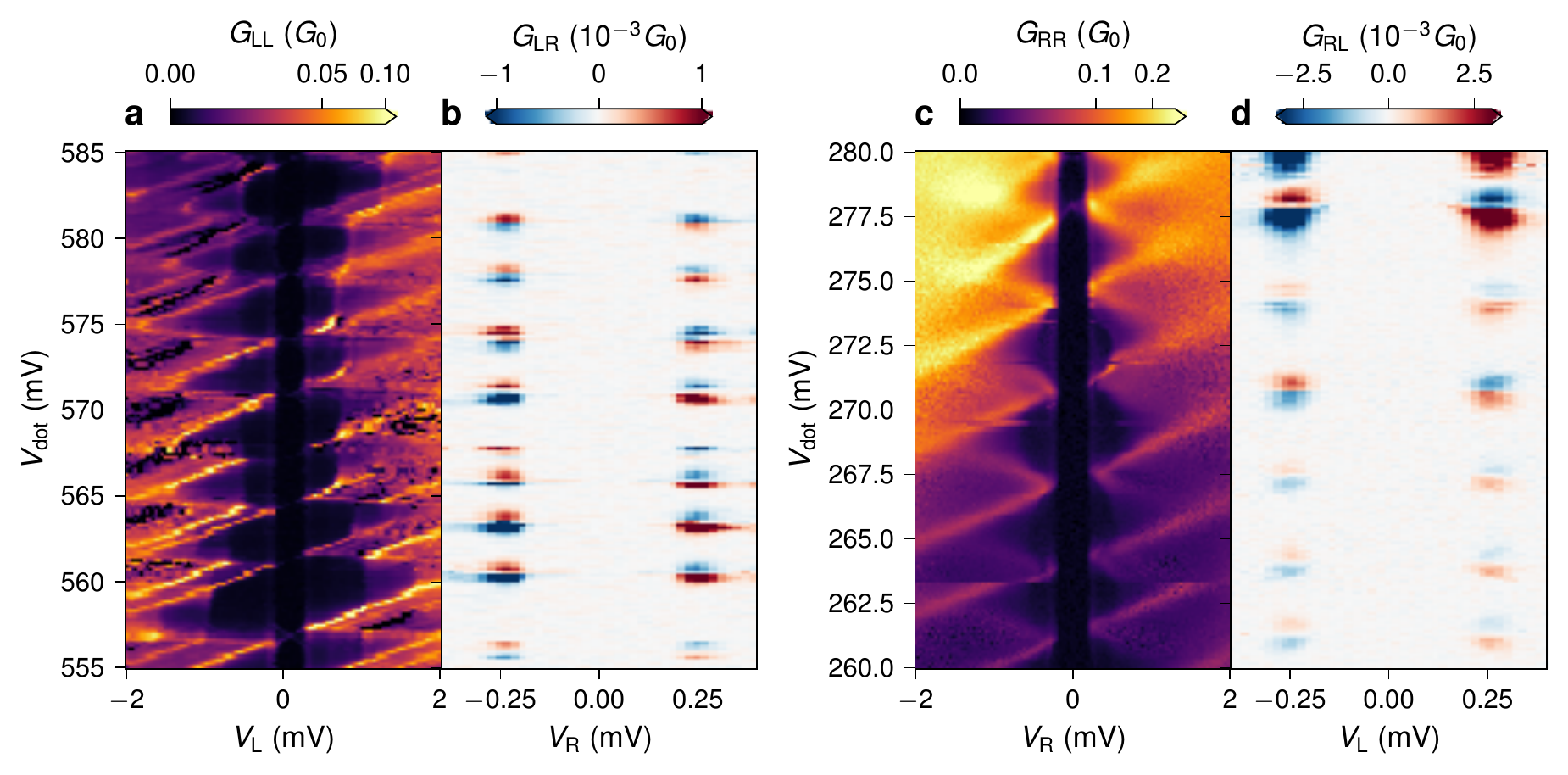}
    \caption{   \textbf{Additional examples of charge filtration} 
    \textbf{a. b.} The local and nonlocal conductance, $G_\mathrm{LL}$ and $G_\mathrm{RL}$, when Device~A is configured to have a QD in the \textit{left} junction and a tunnel barrier on the right, through a series of multiple QD resonances. 
    The structure appearing on the right side of the same device (Figure~2 of the main text) clearly repeats here for every resonance. 
    \textbf{c,~d.} Local and nonlocal conductance of Device~B, respectively, following the same trend as panels~a,~b, again illustrating the generality of the physics.  }
    \label{fig: additional-dots}
\end{figure}

\begin{figure}[h!]
    \centering
    \includegraphics[width=0.4\textwidth]{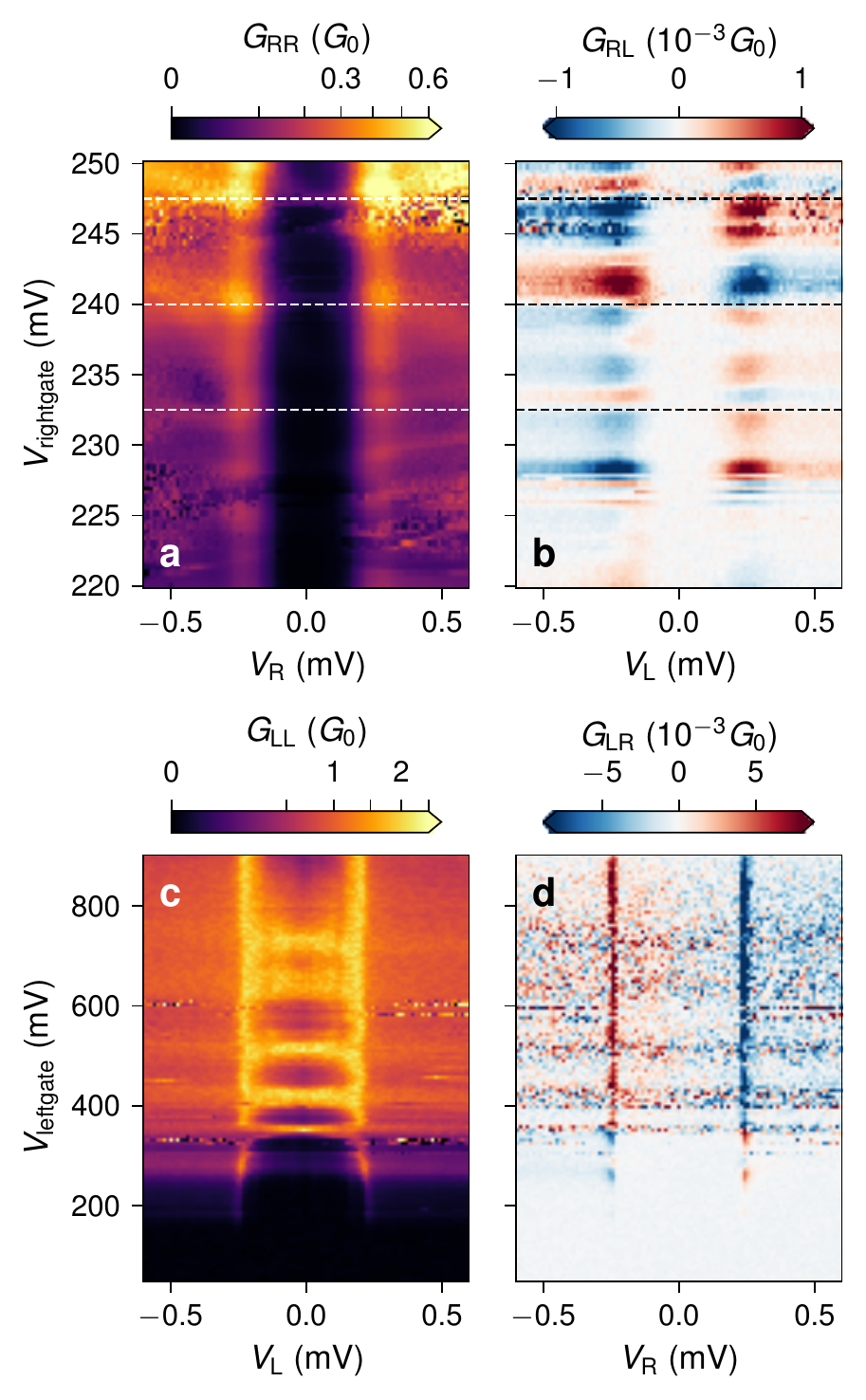}
    \caption{\textbf{Clean vs. disordered junctions} The analysis given in the main text focuses only on the dot's role of introducing a sharply peaked and well-controlled energy dependence into the tunneling matrix elements between the hybrid nanowire and the right lead. 
    The origin of that energy dependence, such as large level spacing or finite charging energy of the dot, does not factor into the discussion. 
    Therefore, we suggest that this discussion can be extended to tunnel junctions without intentionally defined QDs, where disorder or barrier bending can generate the required energy dependence. 
    Such behaviors can be commonly observed in typical nanowire devices \cite{Puglia2020,Danon2020}.
    To illustrate this wide applicability of our model, we perform similar measurements to those in Figure~2 of the main text using three-terminal NSN setups involving only tunnel junctions on each side \textit{without} QDs.
    Here the gate controlling one of the tunnel junctions is varied, serving a similar role to $V_\mathrm{dot}$ in the main text.
    \textbf{a, b.} The local and nonlocal conductance ($G_\mathrm{RR}$ and $G_\mathrm{LR}$) measured on the left junction of the Device~A, while sweeping the left gate. 
    Both sides of the device are configured to be tunnel junctions by setting the outer finger gates to high voltages, and only the inner gates are used to define tunnel barriers.
    As discussed above, the local conductance exhibits typical disordered-junction behavior: a superconducting gap on top of local, gate-dependent resonances. 
    The energy dependence of the tunneling matrix element stems mainly from these resonances. 
    A resonance crossing zero energy changes the tunneling preference from in favor of one type of charge carrier to the other. 
    The nonlocal conductance indicates these crossings by a change of its sign, in complete analogy to the observation in Figure~2 of the main text.  
    Guides to the eye indicate where local resonances and the global nonlocal conductance phase change coincide. 
    \textbf{c,~d.} Local and nonlocal conductance measured on a clean tunnel junction, namely the left junction of Device~C. 
    This device was fabricated similarly to Devices~A and B. 
    However, the tunnel junctions separating the nanowire and the leads were kept to a short distance of \SI{50}{nm} without multiple finger gates complicating the potential landscape. 
    The local conductance $G_\mathrm{LL}$ of this junction exhibits both quantized conductance plateaus and Andreev enhancement~\cite{Heedt2020}, indicative of its high quality. 
    The nonlocal conductance $G_\mathrm{LR}$ is anti-symmetric and experiences a small number of sign flips over an extended tunnel gate range.  Strikingly, the global sign of the NLC does not change above $V_\mathrm{left gate} = \SI{0.4}{V}$. 
    Careful inspection reveals that the three sign flips in this measurement all coincide with a level crossing in the local signal, further corroborating our model.}
    \label{fig:ballistic}
\end{figure}

\begin{figure}
    \centering
    \includegraphics[width=\textwidth]{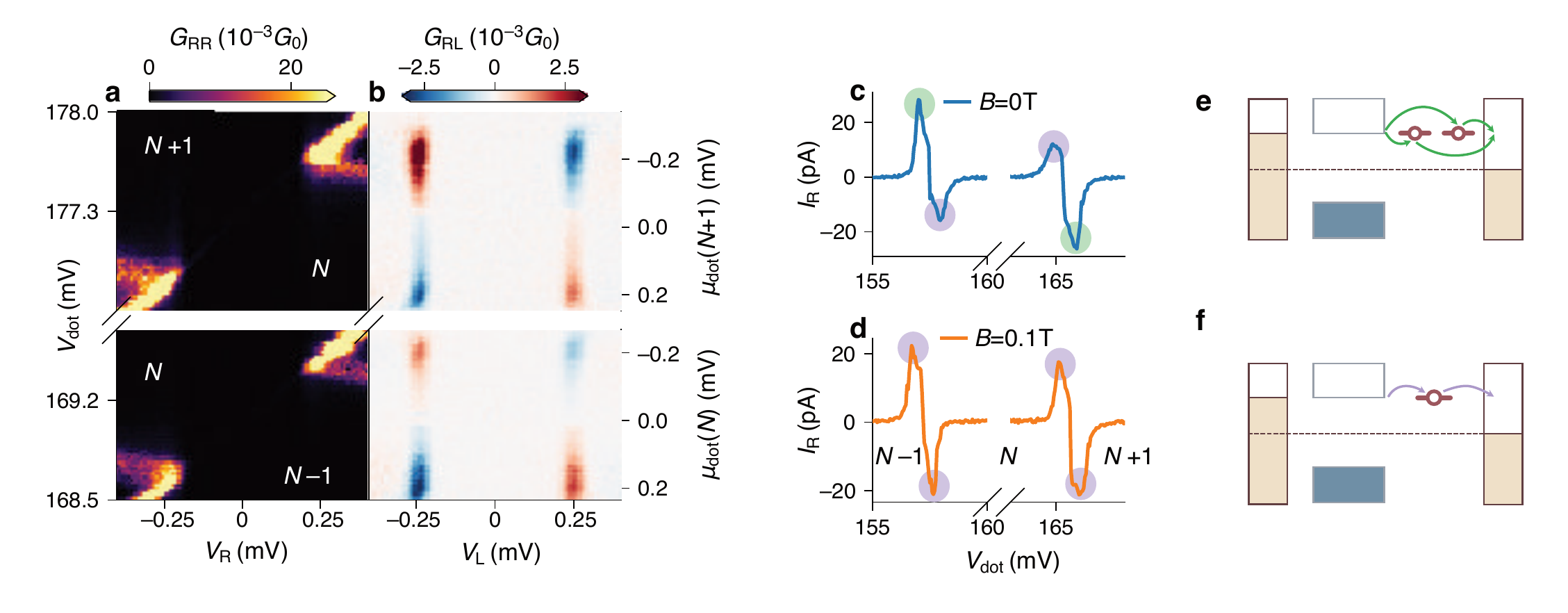}
    \caption{\textbf{Lifting of spin degeneracy}
    \quad In the main text, we stated that the unequal amplitude of $G_\mathrm{RL}$ peaks under different $V_\mathrm{dot}$ in Figure~2 is due to Kramer's degeneracy. 
    Here we provide more details to support this claim.
    \textbf{a. b.} Local (a) and nonlocal (b) conductance at $B=0$, similar to Figure~2(e)(f), through both the $N-1 \rightarrow N$ and $N\rightarrow N+1$ QD resonances shown in Figure~2b.
    The $G_\mathrm{RL}$ peaks on the sides of $N+1$ and $N-1$ ground state charge occupancy are larger than those on the side of $N$, where $N$ is odd.
    \textbf{c. d.} The right-side current $I_\mathrm{R}$ under finite left-bias injection is measured as a function of $V_\mathrm{dot}$ at $B_\mathrm{X}$ = 0 (c, blue) and \SI{0.1}{T} (d, orange).
    The ground-state charge occupation numbers of the dot are labeled in d for each $V_\mathrm{dot}$ range. 
    In the $B=0$ curve, the two peaks circled in green are taller than the two in purple, reproducing the $G_\mathrm{RL}(V_\mathrm{dot}, V_\mathrm{L}=\pm \SI{250}{\micro V})$ trend in panel~b.
    No such even-odd peak intensity relation is observed in the $B=\SI{0.1}{T}$ curve.
    \textbf{e.} Sketch of transport through the dot for the peaks circled in green.
    These two peaks involve a dot transport cycle beginning with $N-1 \rightarrow N$ and $N+1 \rightarrow N$ charge transitions, respectively.
    Since each orbital is doubly degenerate at $B=0$ and the ground state has an even number of electrons, two possible transport channels are available for each of these transitions.
    \textbf{f.} Sketch of transport through the dot for the peaks circled in purple.
    In the $B=0$ case, these two peaks correspond to the dot beginning the transport cycle with $N$ electrons.
    Since the ground state already has an odd occupation, only one transport channel is available to go to either $N+1$ or $N-1$, resulting in a lower current than when a doubly spin degenerate level is available as the excited charge state.
    When the applied $B$ field produces a Zeeman splitting and lifts Kramer's degeneracy, only one transport channel is available regardless of the ground state charge occupation, eliminating this even-odd effect in $I_\mathrm{R}(V_\mathrm{dot})$.}
    \label{fig: deg-lift}
\end{figure}

\begin{figure}
    \centering
    \includegraphics[width=0.45\textwidth]{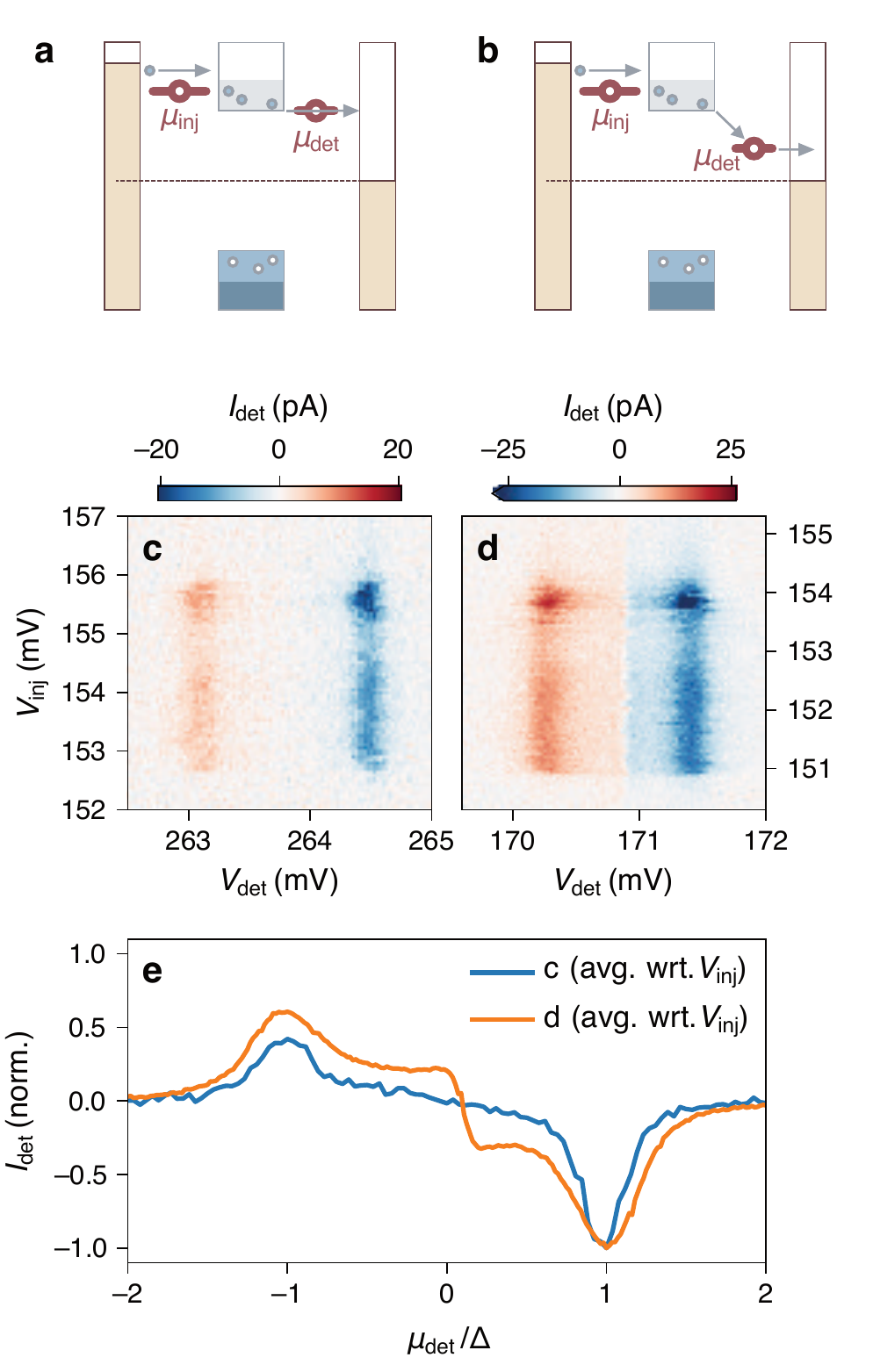}
    \caption{\textbf{Effects of inelastic tunneling through the QD}
    In the main text the energy dependence of the transmission amplitude through the QD was simplified to be $\delta (E-\mu_\mathrm{dot})$.
    However, this is only true in the absence of two effects: inelastic tunneling and elastic co-tunneling.
    Here we discuss the role of inelastic tunneling in the detecting junction in the energy relaxation experiments (figure 5 of the main text).
    \textbf{a.} Energy diagram of the relaxation measurement setup, where only elastic sequential tunneling is allowed in the detector junction.
    Here the transmission amplitude through the detector QD can be regarded as a smoothed delta function.
    \textbf{b.} Energy diagram of the measurement where inelastic tunneling through the detector QD is allowed by increased tunneling rate.
    Inelastic tunneling enables current detection even when the QD energy is lower than $\Delta_\mathrm{ind}$ and no quasiparticle state exists in the hard gap.
    \textbf{c.} Current measured in the detector junction as a function of both injecting and detecting QD voltages in the absence of inelastic tunneling through the detecting QD.
    As the sketch in panel~a shows, no current can be detected when $\mu_\mathrm{dot}$ lies inside the gap.
    \textbf{d.} Similar to panel~c, but with higher cutter gate voltages so that inelastic tunneling through the detector QD becomes appreciable as a result of the higher tunneling rate \cite{Fujisawa1998}.
    In accordance with panel~b, current can be detected when $-\Delta_\mathrm{ind} < \mu_\mathrm{det} < \Delta_\mathrm{ind}$.
    Figure~5 in the main text shows qualitatively the same features.
    Importantly, the presence or absence of inelastic tunneling through the detector QD does not change the conclusion regarding quasiparticle distribution.
    Indeed both panels~c and d show that the upper energy limit of current detection is always the same, namely $\Delta_\mathrm{ind}$, regardless of injector QD energy, verifying that the only quasiparticles at the receiving end of the hybrid S segment are those at the gap edge.
    \textbf{e.} The averaged line scan of $I_\mathrm{det}$ vs.\  $\mu_\mathrm{det}$ from panels~c, d.
    The presence of inelastic tunneling manifests as a plateau-shoulder feature for subgap $\mu_\mathrm{det}$ values.}
    \label{fig: elas-inelas}
\end{figure}

\begin{figure}
    \centering
    \includegraphics[width=\textwidth]{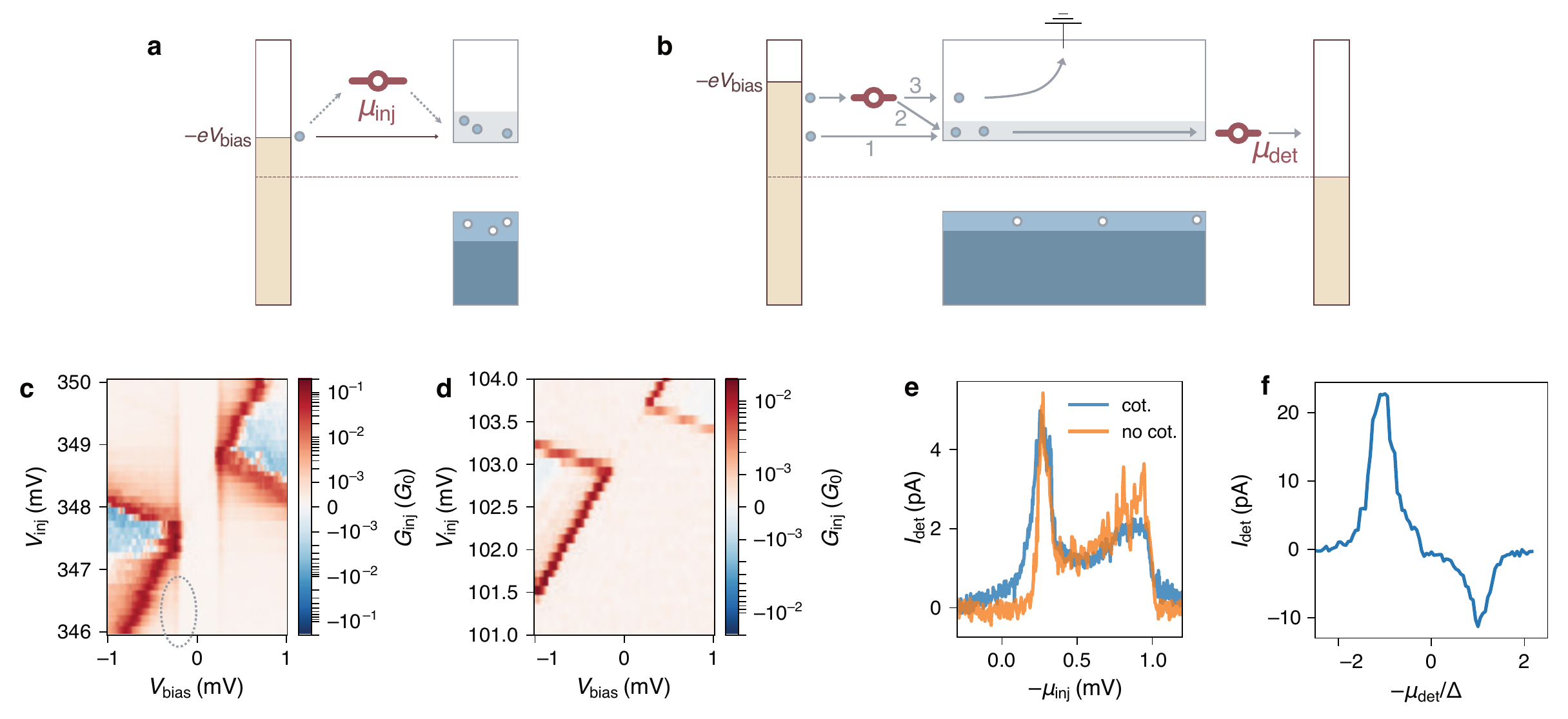}
    \caption{\textbf{Effects of elastic co-tunneling} 
    Here we discuss the effect of elastic co-tunneling through the injector QD and, in combination with inelastic tunneling described in the previous figure, its impact on the interpretation of results presented in Figure~5 of the main text.
    \textbf{a.} Illustration of elastic co-tunneling. 
    When the QD is off-resonance but $E_\mathrm{F}$ in an N lead is aligned with the coherence peak in the S lead, electrons can be directly injected into S via elastic co-tunneling through the QD.
    \textbf{b.} Illustration of an alternative explanation of the absence of high-energy QPs on the receiving side.
    In the main text, we argued that our inability to detect QPs with higher energy than $\Delta_\mathrm{ind}$ is due to energy relaxation inside the hybrid S lead.
    In light of the discussions above regarding both elastic co-tunneling and inelastic tunneling through the QD, an alternative interpretation could be formulated.
    Consider the situation when the injector QD energy is below the left N lead bias but significantly higher than $\Delta_\mathrm{ind}$ in the hybrid S lead.
    This alternative picture starts by considering that electrons coming out of the left N lead can be injected into the high-energy states in S via resonant tunneling through the QD (path 3 in the sketch) as well as directly into the gap edge via elastic co-tunneling (path 1) or inelastic tunneling (path 2).
    Furthermore, this picture assumes the parent gap in the Al film is greater than $\Delta_\mathrm{ind}$ in the InSb wire such that those particles injected via paths 2 and 3 do not have enough energy to enter Al.
    Thus these QPs on the gap edge can only exit the hybrid S lead on the right side through the detector QD, contributing to our measured signal.
    In contrast, since the Al film is grounded, the QPs injected via path one into higher energies can enter Al and drain to the ground before arriving on the right side, thus preventing us from measuring currents that originated from such high-energy injection at all, invalidating the picture of energy relaxation.
    We show in the following panels that this is unlikely to be the case in our experiment because neither path 2 nor 3 contributes significantly to our detected current.
    \textbf{c.} Local conductance measurement through an N-QD-S configuration in which elastic co-tunneling can be observed.
    The conductance feature produced by the process illustrated in panel~a is circled in grey.
    \textbf{d.} Another N-QD-S local conductance measurement where no elastic co-tunneling process is visible due to lower tunnel rates through the barriers.
    Here the vertical co-tunneling lines at $V_\mathrm{bias}=\pm \SI{250}{\micro V}$ are absent.
    \textbf{e.} Comparison of the detected current's dependence on injecting energy when using QDs in panels~c and d as injectors, i.e., \ with and without the presence of co-tunneling.
    Both curves exhibit the same qualitative features: a strong peak when injecting at the gap edge followed by a smoother rise as the injecting energy becomes higher.
    \textbf{f.} Detector current dependence on the detector energy using the QD in panel~d as the \textit{receiver} instead of the injector to characterize the inelastic tunneling through this QD.
    As discussed in Figure~S\ref{fig: elas-inelas}, the inelastic tunneling current contribution compared to resonant tunneling can be read from the plateau-shoulder feature.
    In this case, the plateau is barely visible, and its height is no more than a few percent of the resonant tunneling peak.
    In contrast, in the orange curve in panel~e, when we inject at \SI{1}{mV} energy, the detected current is as high as $\sim60\%$ of the value when injecting at the gap edge.
    This implies that inelastic tunneling across the injector QD cannot be a primary source of contribution to the detected current there.
    Combined with the absence of elastic co-tunneling through this QD, we conclude that most detected QPs were initially injected via resonant tunneling to the high-energy states in the hybrid S lead. Therefore the absence of high-energy QP populations at the detector side must result from complete energy relaxation inside S.
    }
    \label{fig: cot-nocot}
\end{figure}

\begin{figure}
    \centering
    \includegraphics[width=0.5\textwidth]{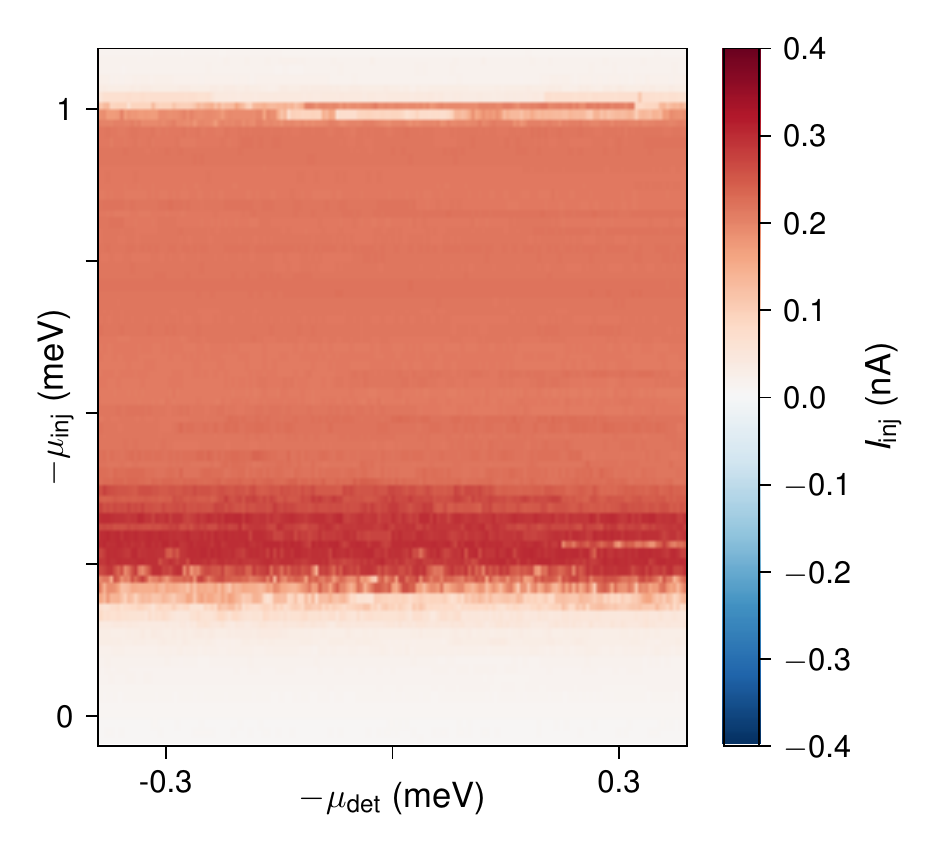}
    \caption{\textbf{Independence of $I_\text{inj}$ on $\mu_\text{det}$}  $I_\text{inj}$ is measured as a function of both $\mu_\text{inj}$ and $\mu_\text{det}$ (see Figure~5 of the main text). 
    $I_\text{inj}$ is independent of $\mu_\text{det}$, ruling out coherent processes such as crossed Andreev reflection or elastic co-tunneling as significant contributors to the transport. }
    \label{fig: Iinj}
\end{figure}

\bibliography{bibliography.bib}